\documentclass{aa}
\usepackage[varg]{txfonts}

\graphicspath{{./}{figures/}}

\usepackage{graphicx,natbib}
\usepackage{hyperref}

\hypersetup{
    colorlinks=true,
    citecolor=blue,
    linkcolor=red,
    urlcolor=black
    }

\def\apjs{ApJS} 
\def\apjl{ApJ} 

\def\aap{A\&A}

\def\ns{n_\mathrm{s}}
\def\nsz{n_{\mathrm{s},0}}
\def\ms{\mu_\mathrm{s}}
\def\vs{v_\mathrm{s}}

\def\ni{n_\mathrm{i}}
\def\mi{\mu_\mathrm{i}}
\def\vi{v_\mathrm{i}}
\def\ve{v_\mathrm{e}}
\def\ne{n_\mathrm{e}}
\def\se{s_\mathrm{e}}
\def\si{s_\mathrm{i}}
\def\ss{s_\mathrm{s}}
\def\me{m_\mathrm{e}}
\def\nh{n_\mathrm{H}}
\def\mh{m_\mathrm{H}}
\def\kis{k_\mathrm{i,s}}
\def\ksi{k_\mathrm{s,i}}
\def\qis{q_\mathrm{is}}
\def\tos{T_{0,\mathrm{s}}}
\def\bs{\beta_\mathrm{s}}
\def\demi{\frac{1}{2}}
\def\svie{\langle\sigma v \rangle_\mathrm{ie}}
\def\svse{\langle\sigma v \rangle_\mathrm{se}}

\begin{document}

\title{Fast methods to track grain coagulation and ionization. II. Extension to thermal ionization}
\titlerunning{}

\author{P. Marchand \inst{1,2}, V. Guillet \inst{3,4}, U. Lebreuilly \inst{5}, M.-M. Mac Low \inst{1}}

\institute{Department of Astrophysics, American Museum of Natural History, Central Park West at 79th Street, New York, NY 10024, USA
\and Institut de Recherche en Astrophysique et Plan\'etologie, Universit\'e Paul Sabatier Toulouse 3, 118 Rte de Narbonne, 31062 Toulouse, France
\and Universit\'e Paris-Saclay, CNRS, Institut d’astrophysique spatiale, 91405, Orsay, France
\and Laboratoire Univers et Particules de Montpellier, Universit\'e de Montpellier, CNRS/IN2P3, CC 72, Place Eug\`ene Bataillon, 34095 Montpellier Cedex 5, France
\and AIM, CEA, CNRS, Université Paris-Saclay, Université Paris Diderot, Sorbonne Paris Cité, 91191 Gif-sur-Yvette, France}

\authorrunning{P. Marchand et~al.}

\date{}

\abstract{Thermal ionization is a critical process at temperatures $T > 10^3$~K, particularly during star formation. An increase in ionization leads to a decrease in nonideal magnetohydrodynamics (MHD) resistivities, which has a significant impact on protoplanetary disks and protostar formation. We developed an extension of the fast computational ionization method presented in our recent paper to include thermal ionization. The model can be used to inexpensively calculate the density of ions and electrons and the electric charge of each size of grains for an arbitrary size distribution. This tool should be particularly useful for the self-consistent calculation of nonideal MHD resistivities in multidimensional simulations, especially of protostellar collapse and protoplanetary disks.}


\maketitle

\section{Introduction}

In \citet[][hereafter paper I]{2021A&A...649A..50M}, we presented a novel method to calculate the coagulation and ionization of grains at a low computational cost. The only ionization source in that model was cosmic rays, which are dominant for isolated dense cores in the interstellar medium. At high temperatures, however, thermal ionization plays a major role in determining the ionization equilibrium of the gas-grain mixture, with a significant impact on the nonideal magnetohydrodynamics (MHD) resistivities.

The first stage of the protostellar collapse is isothermal at $\sim 10$~K until density reaches $\sim 10^{-13}$ g cm$^{-3}$. At this point, the dust-gas mixture becomes opaque to its own thermal radiation and the temperature rises as a core forms and contracts slowly \citep{Larson1969}. At 2000 K, the dissociation of H$_2$ molecules absorbs energy, allowing a rapid second collapse that leads to the creation of the protostar when all H$_2$ is depleted. The thermal ionization of hydrogen occurs during the second collapse, and all hydrogen becomes ionized early in the protostar's life, provoking a drop in resistivities to virtually zero \citep{2016A&A...592A..18M}. Accounting for the thermal ionization of hydrogen is therefore critical to accurately describe the transition between the first core and the protostar, and thus from the nonideal to ideal MHD regime.

Another example of how this could be applicable pertains to chondrule formation. Chondrules are molten grains found in meteorites, whose formation requires rapid heating to $\sim$2000 K \citep{2012M&PSA..75.5387E}. While there is no consensus on this topic, it has been proposed that their creation takes place in magnetic current sheets in protoplanetary disks, which may reach temperatures $>1500$ K \citep{joung2004,mcnally2014}. Those high temperatures would trigger the thermal ionization of K and Na, leading to a sharp decrease in resistivities. Subsequently, the downward gradient in resistivities may create an instability that would allow the magnetic field to pile up in the current sheet \citep{2012ApJ...761...58H}. That phenomenon may be the origin of thunderclaps and extremely localized heating of the grains and gas \citep{2013ApJ...767L...2M}, which are necessary conditions for chrondrule formation. However, readers should refer to \citet{deschturner}, who argue that insufficient alkali metals would evaporate from grains to allow for this instability to act.

In this paper, we focus on extending the ionization model of paper I by including the thermal ionization of one gas-phase species. In Sect.~\ref{SecAnalytical} we analytically derive the grain charge (Sect.~\ref{SecGrainCharge}) and thermal ionization equilibrium (Sect.~\ref{SecEquil}), whose numerical implementations are described in Sect.~\ref{SecNumerical}.  In Sect.~\ref{SecDisc} we discuss applications, and in Sect.~\ref{SecConcl} we present our conclusions.

\section{Analytical method}

\label{SecAnalytical}

\subsection{Grain charge}

\label{SecGrainCharge}
Let us consider the two ionic species $i$ and $s$ of number density $\ni$ and $\ns$. Species $i$ corresponds to all the ions that are exclusively ionized by cosmic rays in the same manner as in paper I, for which we assume an average atomic mass $\mi = m_\mathrm{i} / m_H$. Species $s$, however, undergoes both cosmic-ray and thermal ionization. We define $\nsz$ as the total abundance of species $s$ (both neutral and charged), so that $\nsz$ is an upper bound for $\ns$. We also consider an arbitrary size distribution of dust grains.

Grain charges $Ze$, with $e$ being the electron charge, fluctuate stochastically due to the collection of electrons and the recombination of ions on their surfaces. The grain charge equilibrium \citep[eq. 4.3 of][and eq. 24 of paper I]{DraineSutin} can be written as 
\begin{equation}\label{EqFlux}
  f(Z,\tau_k)(J_i(Z,\tau_k)+J_s(Z,\tau_k))=f(Z+1,\tau_k)J_\mathrm{e}(Z+1,\tau_k),
\end{equation}
where
\begin{equation}
  \tau_k= \frac{a_k k_\mathrm{B}T}{e^2}
\end{equation}
is the reduced temperature of a grain of radius $a_k$ and temperature $T$, while $k_\mathrm{B}$ is the Boltzmann constant. Furthermore, $f(Z,\tau_k)$ is the distribution function of grain charges for a grain of reduced temperature $\tau_k$, and $J_\mathrm{i}$, $J_\mathrm{s}$, and $J_\mathrm{e}$ are the fluxes of species $i$, $s$, and electrons onto the grains, respectively. The fluxes are \citep{DraineSutin}
\begin{equation}
   J_j(Z,\tau_k) = n_j s_j v_j \pi a_k^2 \tilde{J}(Ze/q_j,\tau_k),
\end{equation}
with $n_j$ being the abundance of species $j$ (for $j= i, s$, or $e$), $s_j$ being the sticking probability of species $j$ on grains, $m_j$ and $q_j$ being the mass and charge of species $j$, and $v_j = (8 k_\mathrm{B}T/\pi m_j)^{1/2}$ being the thermal speed of $j$. The polarization factor of grains for species $j$ is $\tilde{J}$, which depends on the relative signs of $Z$ and $q_j$. 

Similarly to the case without thermal ionization, presented in paper I, we need to solve equation (\ref{EqFlux}) for small grains and low temperatures, for which $\tau_k \ll 1$ and $f(Z)$ is only significant for $Z=-1, 0$ and 1; that is, these grains hold a maximum of one charge. Therefore 
\begin{equation}\label{EqNormalization}
f(-1,\tau_k)+f(0,\tau_k)+f(1,\tau_k)=1.
\end{equation}
Equation (\ref{EqFlux}) can be rewritten for Z=0 and Z=-1
as follows:\begin{align}
f(-1,\tau_k) &=f(0,\tau_k)\frac{J_\mathrm{e}(0,\tau_k)}{J_\mathrm{i}(-1,\tau_k)+J_\mathrm{s}(-1,\tau_k)}, \label{EqFm1}\\
f(1,\tau_k) &=f(0,\tau_k)\frac{J_\mathrm{i}(0,\tau_k)+J_\mathrm{s}(0,\tau_k)}{J_\mathrm{e}(1,\tau_k)}. \label{EqFp1}
\end{align}
As in paper I, we have
\begin{align}
\tilde{J}(0,\tau_k) \approx & \left(\pi/ 2\tau_k\right)^{\frac{1}{2}},\\
\tilde{J}(-1,\tau_k) \approx &\; (2 / \tau_k),
\end{align}
where $\tilde{J}(0,\tau_k)$ appears in $J_\mathrm{e}(0,\tau_k)$, $J_\mathrm{i}(0,\tau_k)$, and $J_\mathrm{s}(0,\tau_k)$, while $\tilde{J}(-1,\tau_k)$ appears in $J_\mathrm{e}(1,\tau_k)$, $J_\mathrm{i}(-1,\tau_k)$, and $J_\mathrm{s}(-1,\tau_k)$.
We can then solve the equation system (\ref{EqNormalization})- (\ref{EqFp1}). With $s_\mathrm{i}=s_\mathrm{s}=1$, we obtain
\begin{align}
  f(0,\tau_k) &=  \frac{1}{ 1 + \frac{1}{\alpha_k} \left[\epsilon \Theta +\frac{1}{\epsilon \Theta} \right]},\\
  f(-1,\tau_k) &= \frac{\epsilon^2 \Theta^2}{1 + \alpha_k \epsilon \Theta + \epsilon^2 \Theta^2}, \\
  f(1,\tau_k) &=  \frac{1}{1 + \alpha_k \epsilon \Theta + \epsilon^2 \Theta^2,}
\end{align}
where
$\Theta=s_\mathrm{e} (\mi \mh / \me)^{\frac12}$, $\qis= (\mi / \ms)^\frac12$, and
\begin{equation}
    \epsilon=\frac{\ne}{\ni+\qis\ns}.
\end{equation}

The main difference with paper I is the appearance of the term $\qis$. While $\ni+\ns$ is the total abundance of ions, $\ni +\qis \ns$ is an effective abundance that reflects the relative flux of ions onto grains. The average charge of grains 
\begin{equation}
    Z_k = \sum_{Z=-1}^{Z=1} Zf(Z,\tau_k)
\end{equation}
and the grain-ion recombination enhancement factor 
\begin{equation} 
   \langle \tilde{J}(\tau_k) \rangle = \sum_{Z=-1}^{Z=1} \tilde{J}(Ze/q_\mathrm{i},\tau_k)f(Z,\tau_k)
\end{equation}
thus are given by the same expressions as in paper I,
\begin{align}
    Z_k =& \frac{1-\epsilon^2 \Theta^2}{1+\alpha_k \epsilon \Theta+\epsilon^2 \Theta^2},\label{EqSmallg1}\\
    \langle \tilde{J}(\tau_k) \rangle =& \frac{\frac{2}{\tau_k} (\epsilon^2 \Theta^2 + \epsilon \Theta)}{\epsilon^2 \Theta^2 + \alpha_k \epsilon \Theta +1},\label{EqSmallg2}
\end{align}
where we neglected the recombination of ions on positively charged grains ($\tilde{J}(1,\tau_k)$).

For the larger grains ($\tau_k \gg 1$), the same kind of change needs to be made to the Spitzer equation \citep{Spitzer49,DraineSutin} that governs the grain's electric potential $\psi$. For $\psi<0$, $e^\psi$ represents the repulsion of the flux of electrons by the negatively charged grains, while $1-\psi$ characterizes the attraction of the flux of ions. The flux equilibrium can thus be written as
\begin{equation}
    (\ne \ve \se)e^\psi = (\ni \vi \si + \ns \vs \ss)(1-\psi).
\end{equation}
Introducing the same notations as above, we can write
\begin{equation}\label{EqSpitzer}
    \epsilon = \frac{1-\psi}{\Theta e^\psi},
\end{equation}
which is the same equation as the one-ion model of paper I (eq.\ 34) with the modified expression for $\epsilon$.
The average charge of large grains and the grain-ion recombination enhancement factor yield the same expressions as in paper I,
\begin{align}
    Z_k =& \psi \tau_k,\label{EqLargeg1}\\
    \langle \tilde{J}(\tau_k) \rangle = &  (1-\psi).\label{EqLargeg2}
\end{align}
The average charge and recombination enhancement factor for a mix of small and large grains is assumed to be the sum of the contributions from both equations (\ref{EqSmallg1})+(\ref{EqLargeg1}) and equations (\ref{EqSmallg2})+(\ref{EqLargeg2}) \citep{DraineSutin}.

\subsection{Ionization equilibrium}
\label{SecEquil}
We always assume charge neutrality,
\begin{equation}\label{EqNeutral}
    \ni+\ns - \ne + \sum n_k Z_k = 0.
\end{equation}
This allows us to find the ionization equilibrium for species $i$. In paper I, we considered the balance between the creation of species $i$ by cosmic-ray ionization, and the destruction of species $i$ by recombination with electrons and with grains. In this two-ion model, we also need to consider the charge exchange reactions between species $i$ and $s$. The one-ion model hides and summarizes all the charge transfer reactions between gas-phase species in the choice of $\mi$. Here, we have to explicitly account for the creation of species $i$ by the destruction of species $s$, and vice versa. The ionization equilibrium is then 
\begin{align}\label{EqIonizi}
  & \zeta (\nh-\nsz -\ni) + \ksi (\nh-\nsz -\ni)\ns \nonumber\\
  &= \svie \ne \ni + \ni \vi \sum n_k \pi a_k^2 J_k + \kis (\nsz-\ns) \ni,
\end{align}
where $\zeta$ is the cosmic-ray ionization rate, $\ksi$ and $\kis$ are the chemical reaction rates of species $s \rightarrow i$ and $i \rightarrow s$, respectively, and we consider the recombination rate of ions $i$ with electrons to be $\svie = 2\times 10^{-7} (T/300)^{\frac{1}{2}}$ cm$^3$ s$^{-1}$, based on the recombination rate of HCO$^+$ taken from the UMIST database \citep{McElroy}.
The terms of the form $k_{a,b} n_a n_c$ are the transformation of species $c$ to species $b$, through the chemical reaction with species $a$. Hence the term $(\nh-\nsz-\ni)$ represents the total abundance of neutral species that can be ionized into ion $i$, as $(\nsz-\ns)$ is the total abundance of neutral species $s$ that can be ionized to ion $s$.

The ionization equilibrium for species $s$ is similar, with the addition of a thermal ionization term \citep[][also see Section \ref{SecDisc}]{PneumanMitchell}
\renewcommand{\arraystretch}{0.3}
\begin{equation}
  \left(\frac{d\ns}{dt}\right)_{\begin{array}{l} \th{\scriptstyle thermal}\\ \th{\scriptstyle ionization} \end{array}} = \bs \nh T^{1/2} e^{-\tos/T} (\nsz - \ns),
\end{equation}
with the values of $\beta$ and $T_0$ depending on the species. Table \ref{TableSpecies} summarizes the values of constants specific to species $s$ for the cases of sodium, potassium, and hydrogen.
The ionization equilibrium equation is then
\begin{align}\label{EqIonizs}
  &\zeta (\nsz-\ns) + \left(\frac{d\ns}{dt}\right)_{\begin{array}{l} \th{\scriptstyle thermal}\\ \th{\scriptstyle ionization} \end{array}}  + \kis(\nsz-\ns) \ni \nonumber\\
  &= \svse \ne \ns + \ns \vs \sum n_k \pi a_k^2 J_{k}+ \ksi(\nh-\nsz -\ni)\ns.
\end{align}

Those equations are valid if the Saha equation is valid as well, meaning that there should be a large number of particles within a Debye length of each other. The validity condition is then
\begin{equation}
\frac{4}{3}\pi \nh \lambda_\mathrm{D}^3 \gg 1,
\end{equation}
with
\begin{equation}
\lambda_\mathrm{D}=\sqrt{\frac{k_\mathrm{B}T}{4\pi\ni e^2}}.
\end{equation}

\begin{table*}
  \caption{Constants depending on the choice of species $s$ for sodium, potassium, and hydrogen. The values of $\bs$ and $\tos$ are taken from \citet{PneumanMitchell} (see Section \ref{SecDisc} for an important discussion about those rates), while $\nsz/\nh$ are the same as in \citet{UmebayashiNakano1990}. The values of $\kis$, $\ksi$, and $\svse$ are taken from the UMIST database \citep{McElroy}, by summing the rates over all reactions involving those species in the reduced network of \citet{2016A&A...592A..18M}.}
\label{TableSpecies}
\centering
\begin{tabular}{llllllll}
\hline\hline
  Species   &  $\ms$   & $\nsz/\nh$          & $\bs$ (cm$^3$ s$^{-1}$ K$^{-\frac{1}{2}}$)   & $\tos$ (K)                & $\kis$ (cm$^{3}$ s$^{-1}$) & $\ksi$ (cm$^{3}$ s$^{-1}$)  & $\svse$ (cm$^{3}$ s$^{-1}$) \\
\hline
  Na        &  $22.99$ & $3.1\times 10^{-9}$  &  $1.4\times 10^{-15}$      & $6.0\times 10^{4}$   & $6.2\times 10^{-9}$         & 0                            & $2.78\times 10^{-12} (T/300)^{-0.68}$\\
  K         &  $39.09$ & $2.2\times 10^{-10}$ &  $6.5\times 10^{-15}$      & $5.1\times 10^{4}$   & $6.2\times 10^{-9}$         & 0                            & $2.78\times 10^{-12} (T/300)^{-0.68}$\\
  H         &  $1.0$ & $1.0$              &  $2.0\times 10^{-10}$      & $15.8\times 10^{4}$   & $3.7\times 10^{-14}$         & $3.8\times 10^{-9}$               & $3.5\times 10^{-12} (T/300)^{-0.75}$\\
\hline
\end{tabular} 
\end{table*}

\section{Numerical implementation and tests}
\label{SecNumerical}

Equations (\ref{EqSpitzer}), (\ref{EqNeutral}), (\ref{EqIonizi}), and (\ref{EqIonizs}) need to be solved for $\psi$, $\epsilon$, $\ni$, and $\ns$. In this section, we discuss the solution for this system with four equations and four unknowns. The system could be reduced to three equations, as equation (\ref{EqSpitzer}) is an explicit expression of $\psi$ as a function of $\epsilon$. That would, however, significantly increase the analytical and numerical complexity of the calculation, and it is unclear whether this would lead to better performances or not.

\subsection{Numerical convergence}\label{SecConvergence}

Although the system of equations is valid for a wide range of physically valid parameters, we need to be cautious to ensure numerical convergence toward the solution, especially at high density and temperature. At high density, $\psi$ converges toward zero by a negative value and becomes very small in an absolute value. At high temperature, $\ns$ overwhelmingly dominates $\ni$ due to the thermal ionization. Therefore, the numerical implementation has to be robust for the cases $|\psi|\ll 1$ and $\ns/\ni \gg 1$.

For this purpose, the four equations must be normalized so that they can be written in the form $1+ x = 0$ to avoid sums of very large or very small numbers. It is therefore necessary to include species $s$ in the normalization of the equations to avoid convergence issues at large temperatures when $\ns$ grows much larger than $\ni$. We therefore normalized equation (\ref{EqNeutral}) by the total number of ions $\ni + \ns$, we used both the cosmic-ray ionization rate and the chemical reaction rate $s\rightarrow i$ to normalize equation (\ref{EqIonizi}), and we included the thermal ionization term in the normalization of equation (\ref{EqIonizs}).

Another issue arises from the average grain charge (\ref{EqSmallg1}). Before the thermal ionization starts to be relevant and $\ni\gg\ns$, $\epsilon$ converges toward $1/\Theta$ as density increases (see Fig. 2 of paper 1). When the difference between $\epsilon$ and $1/\Theta$ becomes close to machine precision, the term $1-(\epsilon \Theta)^2$ reaches a lower bound\footnote{For example, for a machine precision of $10^{-16}$, if $|\epsilon-1/\Theta|<10^{-10}$, then $1-(\epsilon \Theta)^2 > 10^{-6}$.} which prevents the convergence of the charge neutrality (equation \ref{EqNeutral}), as the grain charge fails to decrease. A solution to avoid this issue is to replace $\epsilon$ by $\epsilon' +\epsilon_0$, with $\epsilon_0=1/\Theta$. This substitution has to be made in all the equations. The new variable to find is $\epsilon'$ which converges toward zero instead of $1/\Theta$. The term $1-(\epsilon \Theta)^2$ in equation (\ref{EqSmallg1}) is then mathematically equal to and can be replaced by $-2\epsilon'\Theta -(\epsilon'\Theta)^2$, which avoids the lower-bound issue.

\subsection{The normalized system of equations}

The numerical solution of this system requires its normalization, as discussed in Section \ref{SecConvergence}. We define the function $\mathbf{F}(\psi,\epsilon',\ni,\ns)=(f_1,f_2,f_3,f_4)$, with
\begin{align}
  f_1  &= \frac{1-\psi}{(\epsilon'+\epsilon_0) \Theta e^\psi} -1 = 0,\label{EqF1}\\
  f_2  &= 1 - (\epsilon'+\epsilon_0)\frac{\ni+\qis\ns}{\ni+\ns} +\frac{1}{\ni+\ns} \sum n_k Z_k = 0,\label{EqF2}\\
  f_3  &= 1- \frac{\nsz+\ni}{\nh}- \frac{\svie (\epsilon'+\epsilon_0) \ni(\ni+\qis\ns)}{(\zeta+\ksi \ns) \nh} \nonumber\\
                              &- \frac{\ni \vi \sum n_k \pi a_k^2 \langle \tilde{J}(\tau_k) \rangle}{(\zeta+\ksi \ns)\nh}  -\frac{\kis \ni (\nsz-\ns)}{(\zeta+\ksi \ns)\nh} = 0,\label{EqF3}\\
  f_4  &= \left(1-\frac{\ns}{\nsz}\right)\left(1 + \frac{\kis\ni}{\zeta+\beta T^\demi e^{-\frac{T_0}{T}}\nh}\right) \nonumber\\
                              &-\frac{\svse(\epsilon'+\epsilon_0) \ns(\ni+\qis\ns)}{(\zeta+\beta T^\demi e^{-\frac{T_0}{T}}\nh)\nsz}  \nonumber\\
                              &- \frac{\ns\vs  \sum n_k \pi a_k^2 \langle \tilde{J}(\tau_k) \rangle}{(\zeta+\beta T^\demi e^{-\frac{T_0}{T}}\nh)\nsz} \nonumber\\
                              &- \frac{\ksi \ns (\nh - \nsz -\ni)}{(\zeta+\beta T^\demi e^{-\frac{T_0}{T}}\nh)\nsz}= 0,\label{EqF4}
\end{align}
where
\begin{align}
  Z_k &= \psi \tau_k + \frac{-2\epsilon'\Theta -(\epsilon'\Theta)^2}{1+\alpha_k (\epsilon'+\epsilon_0) \Theta+(\epsilon'+\epsilon_0)^2 \Theta^2},\label{EqZk}\\
  \langle \tilde{J}(\tau_k) \rangle &=  (1-\psi) + \frac{\frac{2}{\tau_k} ((\epsilon'+\epsilon_0)^2 \Theta^2 + (\epsilon'+\epsilon_0) \Theta)}{(\epsilon'+\epsilon_0)^2 \Theta^2 + \alpha_k (\epsilon'+\epsilon_0) \Theta +1},\label{EqJk}\\
  \epsilon &= \frac{\ne}{\ns}=\epsilon' + \epsilon_0,\\
  \epsilon_0&= \frac{1}{\Theta},
\end{align}
and
\begin{equation}
      v_\mathrm{i,s} = \left(\frac{8k_\mathrm{B}T}{\pi\mu_\mathrm{i,s}m_\mathrm{H}}\right)^{1/2}.
\end{equation}
Here, $v_\mathrm{i,s}$ and $\mu_\mathrm{i,s}$ stand for either $\vi$ and $\mi$ or $\vs$ and $\ms$.

\subsection{Solving the system}

Similarly to paper I, we used a Newton-Raphson method to solve the equation system.
Let $\mathbf{X} = (\psi,\epsilon,\ni,\ns)$. Starting from an educated guess $\mathbf{X}_0$, we iterated
\begin{equation} \label{eq:N-R}
  \mathbf{X}_{n+1} = \mathbf{X}_n - \mathbb{J}(\mathbf{X}_n)^{-1} \mathbf{F}(\mathbf{X}_n),
\end{equation}
until $||\mathbf{F}(\mathbf{X}_n)|| < \delta \ll 1$. The matrix $\mathbb{J}$ is the Jacobian of the system defined by $\mathbb{J}_{i,j} = \partial f_i / \partial X_j$. The full analytic components of the Jacobian matrix are given in Appendix~\ref{App:Jacobian}.

For reliable convergence, this iterative solution for the system of equations is best started from as close an estimate as possible. For low density ($\nh < 10^{7}$ cm$^{-3}$) and low temperature (typically 10 K), a good starting point is
\begin{align}
    \psi &= \psi_0,\\
    \epsilon &= 0.9999,\\
    \ni &= 10^{-7} \nh,\\
    \ns &= 10^{-4} \ni,
\end{align}
where $\psi_0$ is the solution of 
\begin{equation}
    \frac{1-\psi}{\Theta e^{\psi}}=1.
\end{equation}
For larger densities and temperatures, we recommend solving the system for a gradual increase in those quantities, using the previous solution as a first estimate. In particular, $\ns$ increases very quickly with density and temperature once the thermal ionization starts, and large leaps may lead to convergence failure.

\subsection{Tests}

We solved the normalized system of equations (\ref{EqF1})--(\ref{EqF4}). For testing purposes, we used typical parameters of star-forming environments to compare with the existing literature, but the reader should keep in mind that a wide range of physically sound parameters is possible.
The density spans $\nh=10^{4}$--$10^{25}$~cm$^{-3}$, starting from the lowest density and increasing  $\nh$ gradually. We assumed the same barotropic equation of state as in \citet{2016A&A...592A..18M} to emulate the rise in temperature during a protostellar collapse
\begin{equation}
    T = T_0 \left(1+\left[\frac{\nh}{n_1}\right]^{0.8}\right)^{\frac{1}{2}}\left(1+\left[\frac{\nh}{n_2}\right]\right)^{-0.3}\left(1+\left[\frac{\nh}{n_3}\right]\right)^{-\frac{1.7}{3}},
\end{equation}
with $T_0=10$ K, $n_1=10^{11}~\mathrm{cm}^{-3}$, $n_2=10^{16}~\mathrm{cm}^{-3}$, and $n_3=10^{21}~\mathrm{cm}^{-3}$. We assumed a nonevolving Mathis, Rumpl, Nordsieck (MRN) grain-size distribution \citep{mathis}, with a slope of $-3.5$ between minimum grain size $a_\mathrm{min}=5$ nm and maximum grain size $a_\mathrm{max}=250$ nm, sampled by 26 bins. The grain bulk density is $\rho=2.9$ g cm$^{-3}$ and the dust-to-gas mass ratio is 1\%, which is typical of the insterstellar medium \citep{1978ApJ...224..132B}.
We set $\zeta=5\times 10^{-17}$ s$^{-1}$ \citep{PadovaniHennebelleGalli2013}, $\se=0.5$ \citep{UmebayashiNakano1990}, and $\mi=25$ \citep[close to the molecular mass of Mg, Fe, or HCO$^+$,][]{2016A&A...592A..18M}. 

\begin{figure}
\begin{center}
\includegraphics[trim=3cm 3cm 2cm 2cm, width=0.45\textwidth]{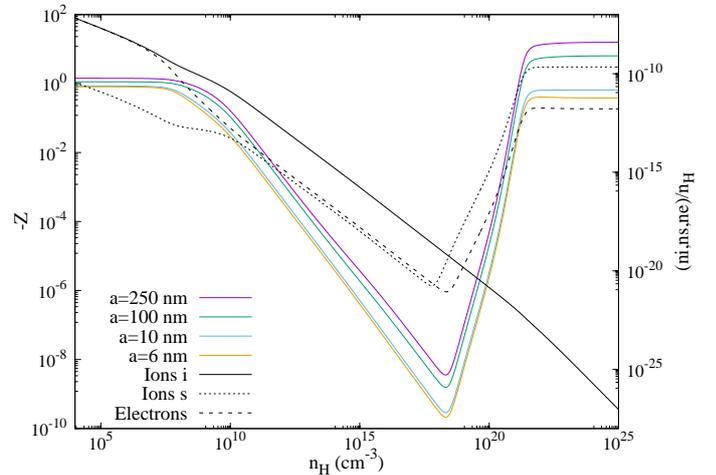}
  \caption{Test of our method for protostellar collapse conditions. Evolution of the average charge of grains (left axis) for several grain sizes (color lines), and the fractional abundance (right axis) of cosmic-ray ionized ions $\ni/\nh$ (solid line), thermally ionized K$^+$ ions $\ns/\nh$ (dotted line), and electrons $\ne/\nh$ (dashed line).}
  \label{FigK}
\end{center}
\end{figure}

\begin{figure}
\begin{center}
\includegraphics[trim=3cm 3cm 2cm 2cm, width=0.45\textwidth]{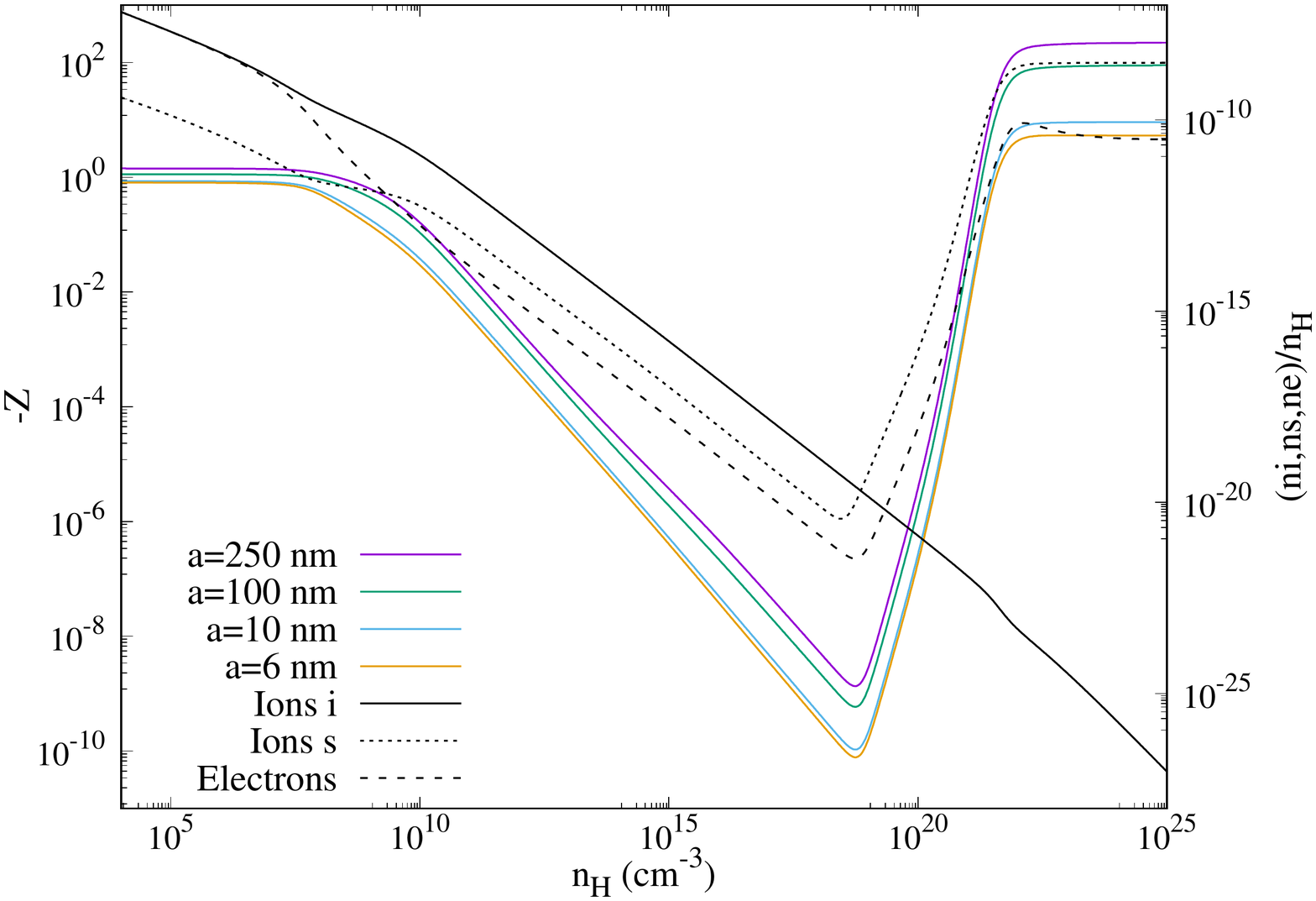}
  \caption{Same as Figure \ref{FigK}, but for Na. The only ions present are the cosmic-ray ionized ones $i$ and Na (ion $s$), which is both cosmic-ray and thermally ionized.}
  \label{FigNa}
\end{center}
\end{figure}

\begin{figure}
\begin{center}
\includegraphics[trim=3cm 3cm 2cm 2cm, width=0.45\textwidth]{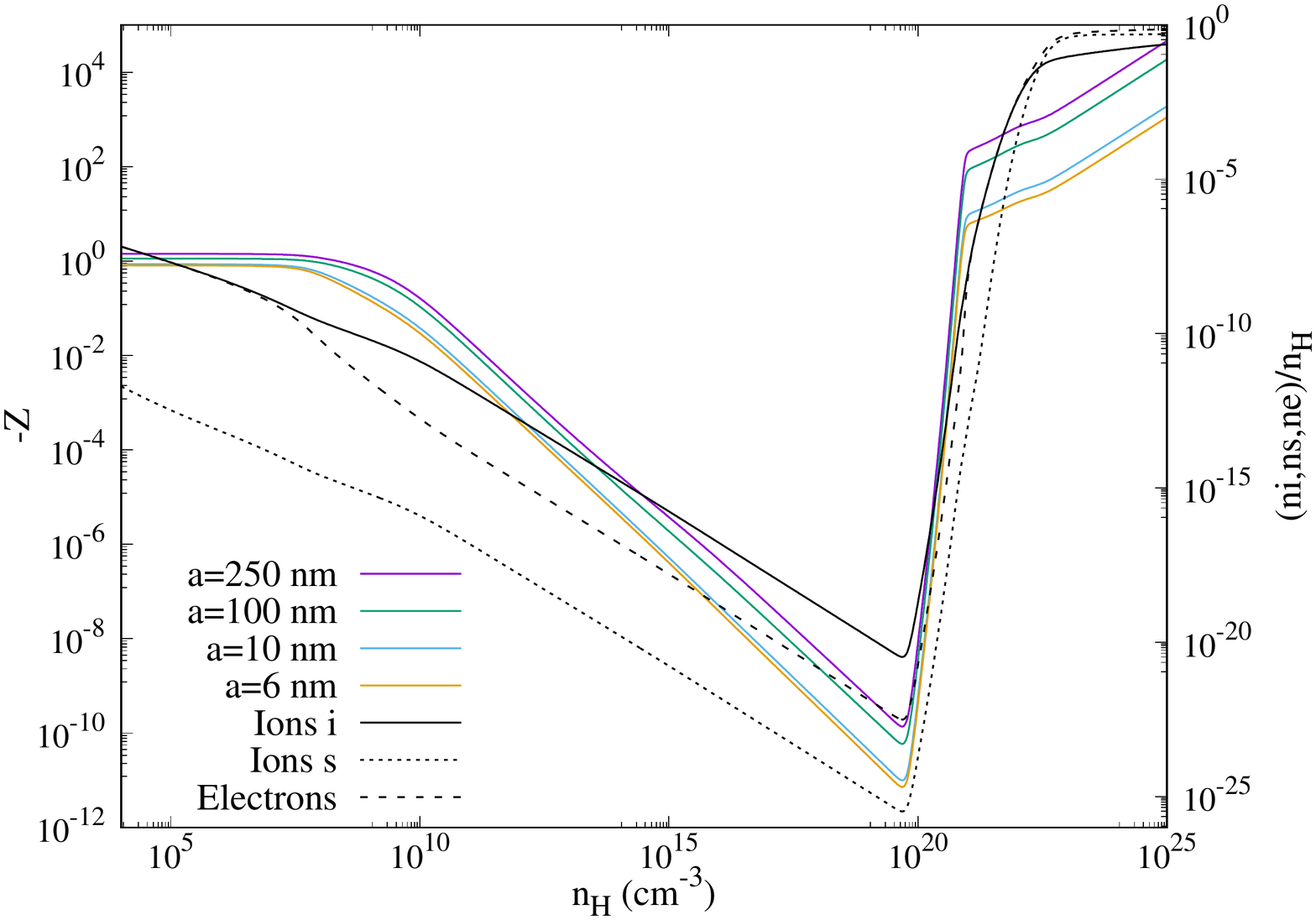}
  \caption{Same as Figure \ref{FigK}, but for H. The only ions presents are the cosmic-ray ionized ones i and H (ions s), which is both cosmic-ray and thermally ionized.}
  \label{FigH}
\end{center}
\end{figure}

In this test, the temperature exceeds several $10^3$~K, above which all grains should be quickly destroyed by evaporation or sputtering \citep{lenzuni}. This is, however, not an issue since at such high temperatures, the contribution of the (computed) charge of grains is negligible compared to that from ions and electrons. It is also possible to combine our method with any grain destruction model. \citet{2016A&A...592A..18M} assumed that the grain evaporation would occur between $\sim 800$ K and $\sim 1600$ K in the density range $\nh \approx 10^{16} - 10^{19}$ cm$^{-3}$, which coincides with the beginning of the thermal ionization of K and Na. 
The species $s$ is assumed to be K, Na, or H, using the values of Table \ref{TableSpecies}. \citet{2015ApJ...811..156D} show that K and Na may originally be confined to grains and have to be evaporated before being available for thermal ionization. Our method is compatible with models accounting for that process since $\nsz$ can be freely modified at any time. For simplicity, however, we assume here that K and Na are already present in the gas phase in quantities given by $\nsz$ in the table.
Here, we present the evolution of $\ni/\nh$, $\ns/\nh$, $\ne/\nh$ and the average grain charge for several bins. The results are displayed in Figures \ref{FigK}, \ref{FigNa}, and \ref{FigH}, respectively.

These figures can be compared to Figure 7 of \citet{2016A&A...592A..18M}. Although the abundances of Na$^+$ and K$^+$ seem overestimated compared to $\ne$, and that of H$^+$ seems underestimated (with a negligible impact on the resistivities), our model reproduces the evolution of abundances of this more detailed calculation at the key points fairly well: K and Na start their ionization around $\nh=10^{18}$ cm$^{-3}$ and $T=1600$ K, and they saturate around $\nh=10^{22}$ cm$^{-3}$ and $T=1.5\times 10^4$ K at the maximum fractional abundance of their respective species.
Furethermore, H has a similar behavior, starting its thermal ionization at $\nh=10^{20}$ cm$^{-3}$ and $T=2650$ K, turning virtually all neutrals into ions by $\nh=10^{23}$ cm$^{-3}$ and $T=7\times 10^4$ K. The main difference with Figure 7 of \citet{2016A&A...592A..18M} is the earlier rise of electron density at $\nh \approx 10^{16}$ cm$^{-3}$ due to the thermionic emission of grains included in the complete chemical calculation. This emission is associated with a drop in neutral grain density and rise in positively and negatively charged grains.

In all three cases, the large input of electrons into the gas significantly increases the grain charges as well. We note that at the hydrogen ionization fraction $n_{\mathrm{H}^+}/\nh > 0.1$ reached
at $\nh>10^{23}$ cm$^{-3}$, the Saha equation is no longer valid, so our model becomes imprecise. However, the resistivities are so low in this regime that ideal MHD is a valid approximation; thus this is never an issue in practice.

\begin{figure}
\begin{center}
\includegraphics[trim=3cm 3cm 2cm 2cm, width=0.45\textwidth]{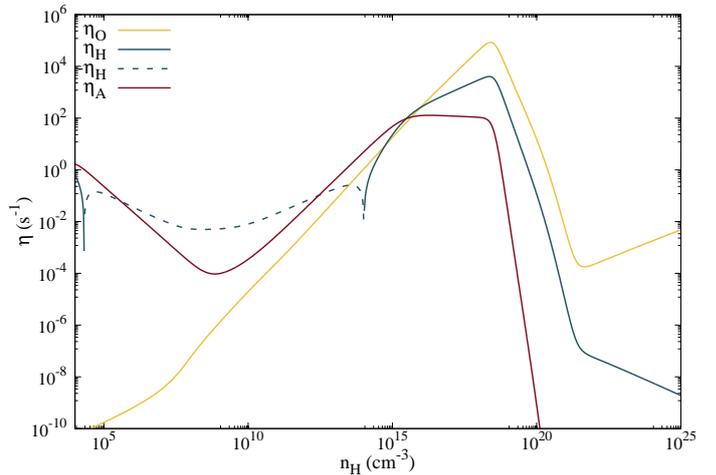}
  \caption{Evolution of the Ohmic (yellow), Hall (blue), and ambipolar (dark red) resistivities for the thermal ionization of K. The dashed line represents the Hall resistivity in negative values. The thin dotted lines are the resistivities in the absence of thermal ionization.}
  \label{FigEtaK}
\end{center}
\end{figure}

\begin{figure}
\begin{center}
\includegraphics[trim=3cm 3cm 2cm 2cm, width=0.45\textwidth]{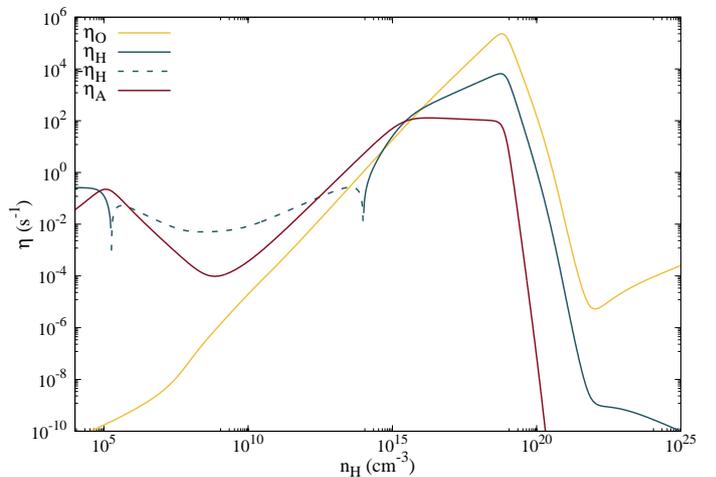}
  \caption{Same as figure \ref{FigEtaK}, but for Na.}
  \label{FigEtaNa}
\end{center}
\end{figure}

\begin{figure}
\begin{center}
\includegraphics[trim=3cm 3cm 2cm 2cm, width=0.45\textwidth]{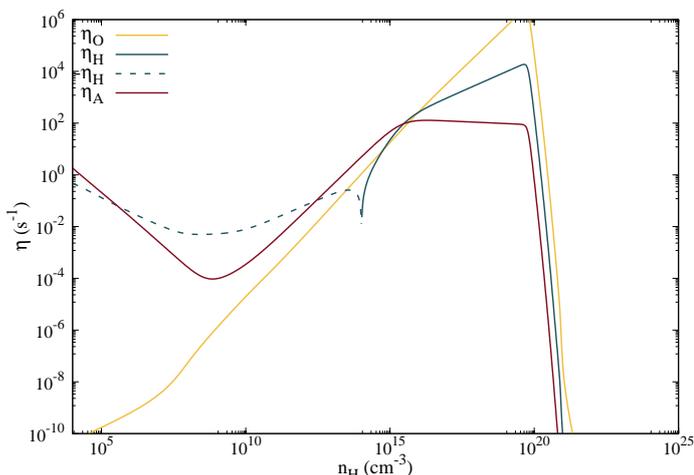}
  \caption{Same as figure \ref{FigEtaK}, but for H.}
  \label{FigEtaH}
\end{center}
\end{figure}

Figures \ref{FigEtaK}, \ref{FigEtaNa}, and \ref{FigEtaH} display the associated nonideal MHD resistivities using the formulae provided by \citet{2016A&A...592A..18M}. For display purposes only, we prescribe a magnetic field \citep{LiKrasnopolskyShang}
\begin{equation}
    B=1.43\times 10^{-7} \mathrm{G}~\nh^{1/2},
\end{equation}
which corresponds to the critical magnetic field strength of a spherical cloud. The magnetic field strength and the resistivities are therefore overestimated compared to protostellar collapse simulations with nonideal MHD for $\nh \gtrsim 10^{12}$ cm$^{-3}$.

The nonideal MHD terms significantly affect the MHD evolution of protostellar collapse and protoplanetary disks for resistivities larger than $\approx 10^{18}$ cm$^{2}$ s$^{-1}$. Nonideal MHD terms are then important at all densities before thermal ionization starts (except the Ohmic diffusion at a low density). At this point, the abundance of charged species in the gas significantly increases, leading to a sharp decrease in all resistivities. The overall behavior is consistent with previous works \citep{Kunz2,2016A&A...592A..18M,2016MNRAS.457.1037W,2019MNRAS.484.2119K}.

\section{Discussion}
\label{SecDisc}
The method presented here is applicable in a wide variety of environments, and it is particularly suited for modeling protostar formation and protoplanetary disks. At later stages of the star formation process, or in the presence of nearby massive stars, photoionization by UV or X-rays may become relevant \citep{2021ApJ...916...32G}. In this case, their ionization rate can simply be added to the cosmic-ray ionization rate $\zeta$ in the system of equations (\ref{EqF1})-(\ref{EqF4}).

We describe our method to calculate the resistivities as fast in comparison to solving a full chemical network. We have implemented the algorithm in the 3D MHD RAMSES code \citep{teyssier}. The code previously calculated the resistivities by interpolating on the precalculated table of \citet{2016A&A...592A..18M}. Without the Hall effect, our thermal ionization algorithm is faster than reading the chemical table, as the calculation needs to be performed only once per cell per time-step. This is different with the Hall effect, which requires, in addition, the self-consistent calculation of resistivities on cell edges \citep{2018A&A...619A..37M}. In this case, the code runs at similar speeds for both methods (it is important to note that this may vary with different implementations). However, the method presented in this paper is much more flexible than a precalculated table because the physical conditions, the chemical composition, and the grain size-distribution can be changed at any point for a self-consistent calculation.

In Table \ref{TableSpecies}, we provide the thermal ionization coefficients for K, Na, and H from \citet{PneumanMitchell}, which is a theoretical work, to match the rates used in \citet{2016A&A...592A..18M}. In the 1960s and 1970s, there were many discussions about the ionization rates of alkali metals. Flame experiments to measure those rates \citep{doi:10.1063/1.1734062,ASHTON197369} resulted in much larger cross sections than theoretically predicted \citep[][see \citealt{combustion1965,doi:10.1063/1.861692} for review]{aller1961abundance,doi:10.1063/1.1670881}. \citet{deschturner} argue that the experimental value of \citet{ASHTON197369} should be preferred to the theoretical value of \citet{PneumanMitchell}. This is debatable as the conditions in flames may be difficult to control and different from astrophysical plasmas (pressure, chemical composition), where Na and K ionize by colliding with H$_2$. The larger coefficient rates would suggest thermal ionization of K and Na at a lower temperature, typically $T=1200$ K instead of $T=1600$ K, which could be of importance in protoplanetary disks. For reference, we provide the experimental values of \citet{ASHTON197369} for K and Na:
\begin{align}
  \beta_\mathrm{K}&=1.0 \times 10^{-8}~ \mathrm{cm}^3~\mathrm{s}^{-1}~\mathrm{K}^{-\frac{1}{2}},\\
  T_{0,\mathrm{K}}&=5.6 \times 10^4~\mathrm{K},\\
  \beta_\mathrm{Na}&=2.0 \times 10^{-9}~ \mathrm{cm}^3~\mathrm{s}^{-1}~\mathrm{K}^{-\frac{1}{2}},\\
  T_{0,\mathrm{Na}}&=5.6 \times 10^4~\mathrm{K}.
\end{align}

\section{Conclusions}
\label{SecConcl}
We detail an extension of the ionization model of \citet{2021A&A...649A..50M} to include the thermal ionization of one gas species. It is possible to increase the number of ionized species by adding their contribution in the same manner, at the price of a higher numerical cost. We have presented the examples of K, Na, and H. Both the chemical abundances and the resistivities show a behavior consistent with previous work. This method is then a powerful tool to self-consistently calculate the ionization of the dust-grain mixture and the nonideal MHD resistivities in hydrodynamical simulations, faster than a complete chemical network. The method is flexible and valid in a wide variety of environments, including star formation and protoplanetary disks. 

\begin{acknowledgements}
We thank the referee for their insightful comments that helped improve the manuscript. We thank J\'er\'emy E. Cohen for his insight on the solving of linear systems.
P. M. acknowledges financial support by the Kathryn
W. Davis Postdoctoral Fellowship of the American Museum of Natural History. U. L. acknowledges financial support from the European Research Council (ERC) via the ERC Synergy Grant ECOGAL (grant 855130). M.-M. M. L. acknowledges partial support from NSF grant AST18-15461.
\end{acknowledgements}

\bibliographystyle{aa}
\bibliography{MaBiblio}

\begin{appendix}
\section{Jacobian components}
\label{App:Jacobian}
The components of the normalized Jacobian matrix required for the solution of equation ~(\ref{eq:N-R}) are
\begin{align}
  \mathbb{J}_{1,1} = \frac{\partial f_1}{\partial \psi} =& \frac{\psi -2}{(\epsilon'+\epsilon_0) \Theta e^{\psi}},\\
  \mathbb{J}_{1,2} = \frac{\partial f_1}{\partial \epsilon'} =& -\frac{1-\psi}{(\epsilon'+\epsilon_0)^2 \Theta e^{\psi}},\\
  \mathbb{J}_{1,3} = \frac{\partial f_1}{\partial \ni} =& 0,\\
  \mathbb{J}_{1,4} = \frac{\partial f_1}{\partial \ns} =& 0,
\end{align}
\begin{align}
  \mathbb{J}_{2,1} = \frac{\partial f_2}{\partial \psi} =& \frac{1}{\ni+\ns}\sum n_k \tau_k,\\
  \mathbb{J}_{2,2} = \frac{\partial f_2}{\partial \epsilon'} =& -\frac{\ni+\qis\ns}{\ni+\ns}+\frac{1}{\ni+\ns}\sum n_k \frac{\partial Z_k}{\partial \epsilon'},\\
  \mathbb{J}_{2,3} = \frac{\partial f_2}{\partial \ni} =& -(\epsilon'+\epsilon_0) \ns\frac{1-\qis}{(\ni+\ns)^2} - \frac{1}{(\ni+\ns)^2} \sum n_k Z_k,\\
  \mathbb{J}_{2,4} = \frac{\partial f_2}{\partial \ns} =& -(\epsilon'+\epsilon_0) \ni\frac{\qis-1}{(\ni+\ns)^2} - \frac{1}{(\ni+\ns)^2} \sum n_k Z_k,
\end{align}
\begin{align}
  \mathbb{J}_{3,1} = \frac{\partial f_3}{\partial \psi} =& \frac{\ni \vi}{(\zeta+\ksi\ns)\nh}  \sum n_k \pi a_k^2,\\
  \mathbb{J}_{3,2} = \frac{\partial f_3}{\partial \epsilon'} =& -\frac{\svie \ni(\ni+\qis\ns)}{(\zeta+\ksi\ns)\nh} \nonumber\\
                                                             &- \frac{\ni \vi}{(\zeta+\ksi\ns)\nh} \sum n_k \pi a_k^2 \frac{\partial J_k}{\partial \epsilon'},\\
  \mathbb{J}_{3,3} = \frac{\partial f_3}{\partial \ni} =& -\frac{1}{\nh} - \frac{\svie (\epsilon'+\epsilon_0) (2\ni+\qis\ns)}{(\zeta+\ksi\ns)\nh} \nonumber\\
                                                        &-\frac{\vi\sum n_k \pi a_k^2 \langle \tilde{J}(\tau_k) \rangle}{(\zeta+\ksi\ns)\nh} - \frac{\kis (\nsz-\ns)}{(\zeta+\ksi\ns)\nh},\\
  \mathbb{J}_{3,4} = \frac{\partial f_3}{\partial \ns} =&  - \frac{\svie(\epsilon'+\epsilon_0)\ni(\qis\zeta-\ksi\ni)}{(\zeta+\ksi\ns)^2 \nh}\nonumber\\
                           &+\frac{\ksi \ni \vi\sum n_k \pi a_k^2 \langle \tilde{J}(\tau_k) \rangle}{(\zeta+\ksi\ns)^2\nh}\nonumber\\
                             &+\frac{\kis\ni(\zeta+\ksi\nsz)}{(\zeta+\ksi\ns)^2\nh}
\end{align}
\begin{align}
  \mathbb{J}_{4,1} = \frac{\partial f_4}{\partial \psi} =& \frac{\ns \vs}{(\zeta+\beta T^\demi e^{-\frac{T_0}{T}}\nh)\nsz} \sum n_k \pi a_k^2,\\
  \mathbb{J}_{4,2} = \frac{\partial f_4}{\partial \epsilon'} =& -\frac{\svse\ns (\ni+\qis\ns)}{(\zeta+\beta T^\demi e^{-\frac{T_0}{T}}\nh)\nsz} \nonumber\\
                                                             &- \frac{\ns \vs}{(\zeta+\beta T^\demi e^{-\frac{T_0}{T}}\nh)\nsz} \sum n_k \pi a_k^2 \frac{\partial J_{k}}{\partial \epsilon'},\\
  \mathbb{J}_{4,3} = \frac{\partial f_4}{\partial \ni} =& \left(1-\frac{\ns}{\nsz}\right)\frac{\kis}{\zeta+\beta T^\demi e^{-\frac{T_0}{T}}\nh} \nonumber\\
                                                        &+ \frac{-\svse(\epsilon'+\epsilon_0) \ns+\ksi \ns}{(\zeta+\beta T^\demi e^{-\frac{T_0}{T}}\nh) \nsz},\\
  \mathbb{J}_{4,4} = \frac{\partial f_4}{\partial \ns} =& -\frac{1}{\nsz}-\frac{\kis\ni+\svse (\epsilon'+\epsilon_0)(\ni+2\qis\ns)}{(\zeta+\beta T^\demi e^{-\frac{T_0}{T}}\nh)\nsz}  \nonumber\\
                                                         &- \frac{\vs \sum n_k \pi a_k^2 \langle \tilde{J}(\tau_k) \rangle}{(\zeta+\beta T^\demi e^{-\frac{T_0}{T}}\nh)\nsz}  - \frac{\ksi (\nh-\nsz-\ni)}{(\zeta+\beta T^\demi e^{-\frac{T_0}{T}}\nh) \nsz},
\end{align}

with

\begin{align}
  \frac{\partial Z_k}{\partial \epsilon'} =& \frac{-(\epsilon'+\epsilon_0)^2 \Theta^3 \alpha_k - 4(\epsilon'+\epsilon_0) \Theta^2 - \alpha_k \Theta}{(1+\alpha_k (\epsilon'+\epsilon_0) \Theta+(\epsilon'+\epsilon_0)^2 \Theta^2)^2},\\
  \frac{\partial J_k}{\partial \epsilon'} =& \frac{2}{\tau_k} \frac{(\alpha_k-1)(\epsilon'+\epsilon_0)^2 \Theta^3 + \Theta + 2(\epsilon'+\epsilon_0) \Theta^2}{(1+\alpha_k (\epsilon'+\epsilon_0) \Theta+(\epsilon'+\epsilon_0)^2 \Theta^2)^2}.
\end{align}

\end{appendix}
\end{document}